\RequirePackage{fix-cm}
\documentclass[pdftex,twocolumn,epjc3]{svjour3}
\usepackage{amssymb}
\usepackage{amsmath}
\usepackage{mhchem}
\usepackage[colorlinks,allcolors=blue]{hyperref}
\usepackage{graphics}
\usepackage{float}
\usepackage[utf8]{inputenc}
\usepackage{indentfirst}
\usepackage{alphabeta}
\usepackage{textcomp}
\usepackage{fancyhdr}
\usepackage{subcaption}
\usepackage{import}
\usepackage{bm}
\usepackage{hyperref}
\usepackage{multirow}
\usepackage{array}
\usepackage{lipsum}
\usepackage{tabularx}
\usepackage{float}
\usepackage{booktabs}
\usepackage{lineno}
\usepackage{chemformula}
\usepackage{mhchem}
\usepackage{hyperref}
\usepackage{setspace}
\usepackage{appendix}
\usepackage{titlesec}

\usepackage{graphicx}
\usepackage{amssymb}
\usepackage{amsmath}
\usepackage{caption}
\usepackage{subcaption}
\usepackage{multirow}
\renewcommand{\sectionmark}[1]{}
\renewcommand{\subsectionmark}[1]{}

\journalname{Eur. Phys. J. C}

\begin{document}

\title{Measurement of the atmospheric $\nu_{\mu}$ flux with six detection units of KM3NeT/ORCA}
\subtitle{The KM3NeT Collaboration\thanksref{a}}

\author{
S.~Aiello\thanksref{1}
\and
A.~Albert\thanksref{2,50}
\and
A.\,R.~Alhebsi\thanksref{3}
\and
M.~Alshamsi\thanksref{4}
\and
S. Alves Garre\thanksref{5}
\and
A. Ambrosone\thanksref{7,6}
\and
F.~Ameli\thanksref{8}
\and
M.~Andre\thanksref{9}
\and
L.~Aphecetche\thanksref{10}
\and
M. Ardid\thanksref{11}
% ...
}
% ----- End author list
% ----- Start address list
\institute{
\label{1}INFN, Sezione di Catania, (INFN-CT) Via Santa Sofia 64, Catania, 95123 Italy
\and
\label{2}Universit{\'e}~de~Strasbourg,~CNRS,~IPHC~UMR~7178,~F-67000~Strasbourg,~France
\and
\label{3}Khalifa University of Science and Technology, Department of Physics, PO Box 127788, Abu Dhabi,   United Arab Emirates
\and
\label{4}Aix~Marseille~Univ,~CNRS/IN2P3,~CPPM,~Marseille,~France
\and
\label{5}IFIC - Instituto de F{\'\i}sica Corpuscular (CSIC - Universitat de Val{\`e}ncia), c/Catedr{\'a}tico Jos{\'e} Beltr{\'a}n, 2, 46980 Paterna, Valencia, Spain
\and
\label{6}INFN, Sezione di Napoli, Complesso Universitario di Monte S. Angelo, Via Cintia ed. G, Napoli, 80126 Italy
\and
\label{7}Universit{\`a} di Napoli ``Federico II'', Dip. Scienze Fisiche ``E. Pancini'', Complesso Universitario di Monte S. Angelo, Via Cintia ed. G, Napoli, 80126 Italy
\and
\label{8}INFN, Sezione di Roma, Piazzale Aldo Moro 2, Roma, 00185 Italy
\and
\label{9}Universitat Polit{\`e}cnica de Catalunya, Laboratori d'Aplicacions Bioac{\'u}stiques, Centre Tecnol{\`o}gic de Vilanova i la Geltr{\'u}, Avda. Rambla Exposici{\'o}, s/n, Vilanova i la Geltr{\'u}, 08800 Spain
\and
\label{10}Subatech, IMT Atlantique, IN2P3-CNRS, Nantes Universit{\'e}, 4 rue Alfred Kastler - La Chantrerie, Nantes, BP 20722 44307 France
\and
\label{11}Universitat Polit{\`e}cnica de Val{\`e}ncia, Instituto de Investigaci{\'o}n para la Gesti{\'o}n Integrada de las Zonas Costeras, C/ Paranimf, 1, Gandia, 46730 Spain
% ...
\and
\label{50}Max-Planck-Institut~f{\"u}r~Radioastronomie,~Auf~dem H{\"u}gel~69,~53121~Bonn,~Germany
% ...
}
% ----- End affiliation list
% ----- End automatically generated KM3NeT info
\thankstext{a}{ km3net-pc@km3net.de}
\thankstext{b}{ dstavropoulos@inp.demokritos.gr (corresponding author)}

\date{\today}

\onecolumn

\maketitle

\twocolumn   

\abstract{
A measurement of the atmospheric $\nu_{\mu}+\bar{\nu}_{\mu}$ flux with energies between 1--100 GeV is presented. The measurement has been performed using data taken with the first six detection units of the KM3NeT/ORCA detector, referred to as ORCA6. The data were collected between January 2020 and November 2021 and correspond to 510 days of livetime, with a total exposure of 433 kton$\cdot$years. Using machine learning classification, 3894 neutrino candidate events have been selected with an atmospheric muon contamination of less than 1$\%$. The atmospheric $\nu_{\mu}+\bar{\nu}_{\mu}$ energy spectrum is derived using an unfolding procedure and the impact of systematic uncertainties is estimated. The atmospheric $\nu_{\mu}+\bar{\nu}_{\mu}$ flux measured using the ORCA6 configuration is in agreement with the values measured by other experiments.}

\section{Introduction} \label{sec1}
Atmospheric neutrinos originate from extensive air showers (EAS), produced when primary cosmic rays interact with the nuclei of the atmosphere \cite{gaisserbook}. The energy spectrum of atmospheric neutrinos covers a wide range, from about $100$ MeV to PeV energies. They are mainly produced in the decays of charged $\pi$ and K mesons --- constitutes as the \textit{conventional} component of the flux:
\begin{equation} \label{eq:pi2munu} 
\pi^{+(-)} \rightarrow \mu^{+(-)} + \nu_\mu(\bar{\nu}_\mu) \text{  ,     } \text{K}^{+(-)} \rightarrow \mu^{+(-)} + \nu_\mu(\bar{\nu}_\mu)
\end{equation}
with muons decaying into
\begin{equation}\label{eq:mu2enunu}
 \mu^{-(+)} \rightarrow e^{-(+)} + \bar{\nu}_e(\nu_e) + \nu_\mu(\bar{\nu}_{\mu})  \, .
\end{equation}

At higher energies, neutrinos that are produced in decays of charmed mesons (mainly D mesons) constitute the \textit{prompt} atmospheric neutrino flux. Due to their longer lifetime ($\tau_\pi \sim 10^{-8} \, \text{s}$) compared to that of D mesons ($\tau_D \sim 10^{-12} - 10^{-13} \, \text{s}$) \cite{pdg}, charged $\pi$ and K mesons travel larger distances before decaying and experience higher energy losses. Conventional neutrinos have a softer energy spectrum than prompt neutrinos. Simulations indicate that for vertically incoming neutrinos, the crossover between conventional and prompt fluxes occurs in the energy range of $10^5$–$10^6$ GeV, and that the crossover energy increases with zenith angle \cite{prompt}. As a product of cosmic ray collisions in the atmosphere, the flux of atmospheric neutrinos approximately follows a power law, $dN/dE \propto E^{\gamma} $.\par 
As a result of Eqs. \ref{eq:pi2munu} and \ref{eq:mu2enunu}, the ratio of $\nu_{\mu}+\bar{\nu}_{\mu}$ to $\nu_{e}+\bar{\nu}_{e}$ in the conventional flux is about 2, increasing with energy, as higher energy muons can reach the Earth before decaying. The flux of neutrinos in the vertical direction is lower than the flux near the horizon, due to the different path-lengths of the parent particles in the atmosphere before decaying \cite{honda2002}. The effect of the Earth's magnetic field on the primary cosmic rays can be neglected for neutrinos above a few GeV \cite{asymetry}.\par 
The $\nu_{e}+\bar{\nu}_{e}$ flux has been measured by Fréjus \cite{frejus}, IceCube \cite{icecubenue1,icecubenue2}, Super-Kamiokande \cite{sk} and ANTARES \cite{antares2}, while the atmospheric $\nu_{\mu}+\bar{\nu}_{\mu}$ flux has been measured by Frejus \cite{frejus}, AMANDA \cite{amanda}, Super-Kamiokande \cite{sk}, IceCube \cite{icecube1,icecube2,icecube3}, and ANTARES \cite{antares1,antares2}. MACRO studied the $\nu_{\mu}+\bar{\nu}_{\mu}$ flux to perform neutrino oscillation studies \cite{macro}. In ref. \cite{icecube3}, astrophysical neutrinos were also included in the measurement by IceCube. Moreover, IceCube has measured the seasonal variation of the atmospheric $\nu_{\mu}+\bar{\nu}_{\mu}$ flux \cite{icseasvar}. While above 100 GeV large neutrino telescopes are sensitive and several different measurements of the the atmospheric $\nu_{\mu}+\bar{\nu}_{\mu}$ flux have been performed, limited experimental information exists in the region below. The measurement by Frejus do not account for neutrino oscillations, which had not been discovered at the time. The Super-Kamiokande experiment provides precision measurements up to 10 GeV, since the experiment has been designed for neutrino energies between 100 MeV and few GeV. Consequently, the energy region between 10 GeV and 100 GeV would benefit from additional experimental data. \par

The paper is structured as follows. In Section \ref{sec2}, the KM3NeT/ORCA detector is introduced. In Section \ref{sec3}, the data and Monte Carlo (MC) simulation used in the analysis are presented. In Section \ref{sec4}, the procedure to select a high-purity atmospheric neutrino event sample from data collected with a partial configuration of the detector is described. In Section \ref{sec5}, the $\nu_\mu + \bar{\nu}_\mu$ charged-current (CC) energy spectrum is obtained, exploiting an \textit{unfolding} procedure. In Section \ref{sec6}, the studies to evaluate the impact of systematic uncertainties on the measurement are presented. Finally, the results are shown in Section \ref{sec7} and discussed in Section \ref{sec8}.

\section{The KM3NeT/ORCA detector} \label{sec2}
The KM3NeT Collaboration is currently constructing two detectors with different scientific goals in the Mediterranean Sea \cite{loi}. The ARCA (Astroparticle Research with Cosmics in the Abyss) detector is located about 80 km offshore Sicily, Italy, at a depth of 3450 m. The primary goal of ARCA is the detection of high-energy astrophysical neutrinos. The ORCA (Oscillation Research with Cosmics in the Abyss) detector is located 40 km offshore Toulon, France, at a depth of 2450 m. The goal of ORCA is to study atmospheric neutrino oscillations in the energy range of a few GeV, and to determine the neutrino mass ordering \cite{orcanom}. The ARCA and ORCA detectors are built using the same technology. \par 

The ORCA detector is an array of photomultiplier tubes (PMTs) for the detection of Cherenkov radiation emitted along the path of neutrino-induced particles in seawater. The 3-inch Hamamatsu PMTs \cite{pmts} are packed in groups of 31 into high-resistance glass spheres in order to endure the pressure at the sea depths. The spheres contain also the required electronics for the operation of the PMTs and the data transmission \cite{electronics}, as well as instruments for the orientation \cite{clb}, position and time calibration \cite{nanobeacon}. These assemblies called Digital Optical Modules (DOMs) \cite{dom_paper} and provide an almost $4\pi$ solid angle coverage.\par
The DOMs are evenly distributed across vertical support structures, the Detection Units (DUs) \cite{ppmdu}. Each DU hosts 18 DOMs, is anchored to the seabed and is kept taut and aloft by the buoyancy of the DOMs and of a buoy which is tied to its top. The distance between two consecutive DOMs in ORCA is about 9 m, and the horizontal separation between the DUs is about 20 m. This geometry was optimised to enhance the sensitivity to GeV neutrinos. When completed, the ORCA detector will consist of 115 DUs, instrumenting a mass of about 7 Mton of seawater. \par 
When the voltage amplitude at the PMT anode exceeds a predefined threshold, a \textit{hit} is recorded and a set of information containing the duration of the pulse over the threshold, the time of the pulse leading edge, and the PMT identifier, are registered and transmitted to shore. On-shore, the data are processed and filtered by a computer farm using dedicated trigger algorithms and the resulting events are stored on disk. \par 
Events recorded in a Cherenkov neutrino telescope have different signatures depending on the physics processes that are involved. Atmospheric muons, residuals of cosmic ray showers, produce long down-going tracks with an abundant flux reaching the detector. Muons mainly from CC interactions of muon neutrinos are selected by applying geometrical cuts - i.e. selecting upgoing track-like events. The electron and tau neutrino CC interactions and the neutral-current (NC) interactions of all neutrino flavours, produce mainly shower-like events, characterised by a more localised area of light emission. 

\section{Data and MC simulation} \label{sec3}

The data used in this analysis have been collected with the 6-DU configuration of the ORCA detector. The data stream is organised in periods of data taking, referred to as runs, with a duration of about 6 hours. Runs for which more than half of the PMTs exceed their maximum measurable hit rate as a result of extreme environmental conditions, such as bioluminescence, are not considered in the analysis. Moreover, a fraction of the operational time was devoted to calibration and test periods of data-taking. The corresponding data are removed from the analysis. The livetime of the selected dataset is 510 days, with an equivalent exposure of 433 kton$\cdot$years \cite{orca6_osci}. \par
A run-by-run Monte Carlo simulation strategy \cite{antares_rbr} has been followed, in which each run is simulated in order to account for the variation of the detector data-taking conditions. \textit{In-situ} calibration constants related to the time offsets and response of PMTs are regularly extracted and applied to data and MC \cite{k40calib}.\par 
Atmospheric neutrinos have been generated using the gSeaGen software \cite{gseagen}, which is based on the GENIE \cite{genie} neutrino generator (both neutrinos and anti-neutrinos are mentioned as neutrinos hereafter, since ORCA can not discriminate between them). The CC neutrino interactions have been simulated for all neutrino flavours, while the NC interactions have been simulated for muon neutrinos and scaled to account also for the other flavours. Event weights have been assigned to account for the atmospheric neutrino flux and neutrino oscillation probabilities. The HKKM14 conventional atmospheric neutrino flux model has been used \cite{hkkm14}. Neutrino oscillation probabilities have been computed with OscProb \cite{oscprob}, assuming the mass eigenstates follow a normal ordering, and using the oscillation parameters from NuFIT 5.2 \cite{nufit} global oscillation fits. Atmospheric muons have been simulated with the MUPAGE software \cite{mupage,mupageparam}. \par 
The GEANT4-based \cite{geant4} KM3Sim \cite{km3sim}, a custom software package, is used for the light propagation of atmospheric neutrino events. In KM3Sim, a full simulation of photons is performed, by tracking in detail the individual secondary particles that are produced at the generation level, as well as the emitted Cherenkov photons. For higher neutrino energies ($E_{\nu}>500$ GeV) and for atmospheric muons, an internal KM3NeT software package is used for the simulation of the light production and propagation, which evaluates the number of photoelectrons recorded by a PMT using tabulated probability distribution functions (PDFs) of the photon arrival time. \par 
Noise hits are simulated on the basis of the measured PMT rates from data. The PMT signals of the simulated hits are digitised. Finally, the stream of simulated events is processed using the same trigger algorithms used for data. \par 
All data and MC simulated events are reconstructed using two KM3NeT-built software packages: one assuming a track-like topology \cite{trackreco} and another assuming a shower-like topology \cite{showereco}. 

\section{Event selection} \label{sec4}

\subsection{Pre-selection}

\begin{figure}[h]
\centering
\resizebox{0.5\textwidth}{!}{
  \includegraphics{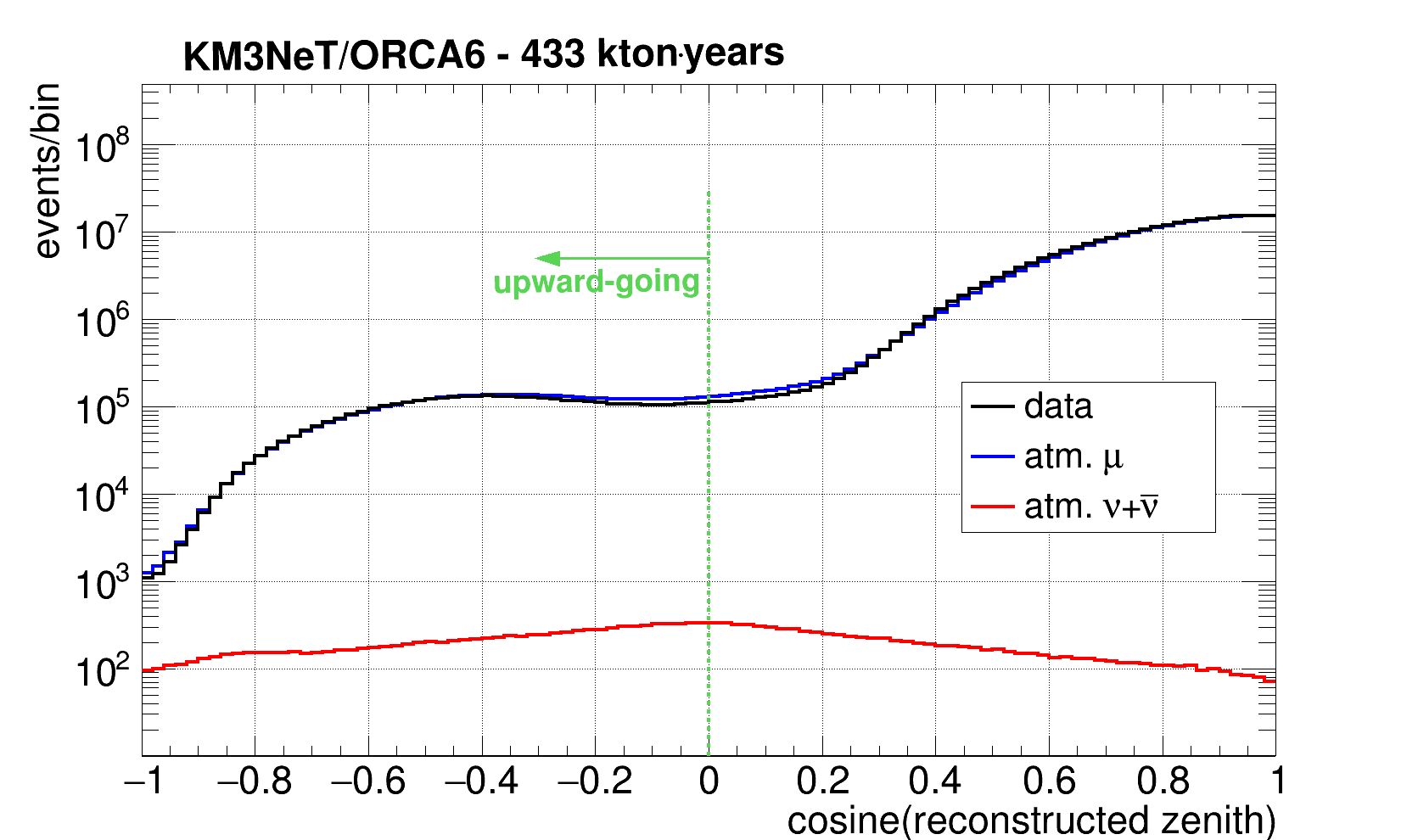}
}
%\vspace{5cm}       % Give the correct figure height in cm
\caption{Distribution of the reconstructed cosine zenith angle for data (black line) and MC simulated events (blue line for atmospheric muons, red line for neutrinos) after applying the anti-noise cuts.}
\label{fig:cos_zen_antinoise}       % Give a unique label
\end{figure}

Pure noise hits (mainly caused by \ce{^{40}K} decays) can produce triggers that lead to poorly reconstructed events. The contribution of pure noise events is suppressed by requiring a minimum number of hits used in the track reconstruction, and a quality cut on the logarithm of the maximum likelihood of the track reconstruction. These criteria are hereafter mentioned as \textit{anti-noise cuts}. The distribution of the cosine of reconstructed zenith angle ($-1$ corresponds to a vertical upward-going direction) for data and MC simulated events is shown in Fig. \ref{fig:cos_zen_antinoise}, after the application of the anti-noise cuts. \par 

The main background source in a neutrino Cherenkov detector comes from atmospheric muons. By selecting only upward-going tracks, an atmospheric muon background rejection of $\sim98\%$ is achieved, while $\sim54\%$ of the atmospheric neutrinos are preserved, with respect to the sample of events after the anti-noise cuts. \par

\subsection{BDT classification}
The event selection procedure is eventually reduced to a binary classification problem: the discrimination between atmospheric neutrino events and atmospheric muon events misreconstructed as upward-going. \par
\begin{figure}[h]
\centering
\resizebox{0.5\textwidth}{!}{
  \includegraphics{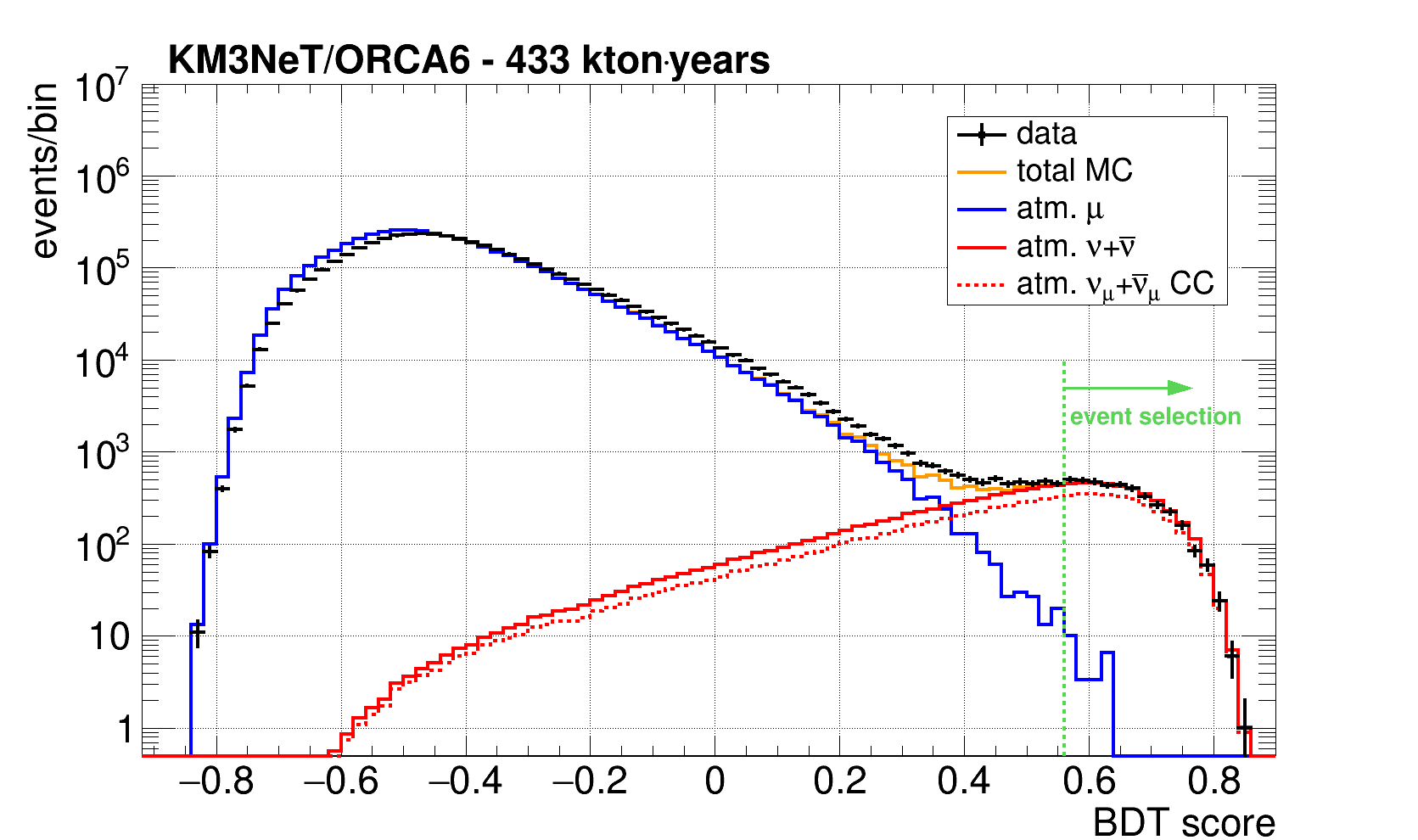}
}
\caption{BDT score distribution for data and MC simulated events fulfilling the pre-selection requirements.}
\label{fig:bdt_score_eval_cand}
\end{figure}

 
For the final event classification an adaptive Boosted Decision Trees (BDT) algorithm is used, implemented in \textit{TMVA} \cite{tmva}. Dedicated MC simulated event samples are produced in order to train the BDT classifier. The training for the signal is performed with a sample of CC atmospheric muon neutrino events. For the background, the BDT is trained on atmospheric muon events. The pre-selection criteria are applied to both training samples. Variables defined on triggered hits and on the basis of the Cherenkov hypothesisfor the origin of the hits, the quality of the event reconstruction, the event topology, as well as the deposited charge in the detector are used as BDT features. The architecture of the BDT classifier has been optimised to obtain the highest classification efficiency. The distribution of the BDT score for the data and MC simulated events fulfilling the pre-selection requirements is presented in Fig. \ref{fig:bdt_score_eval_cand}.  \par

\begin{figure}[h]
\centering
\resizebox{0.5\textwidth}{!}{
  \includegraphics{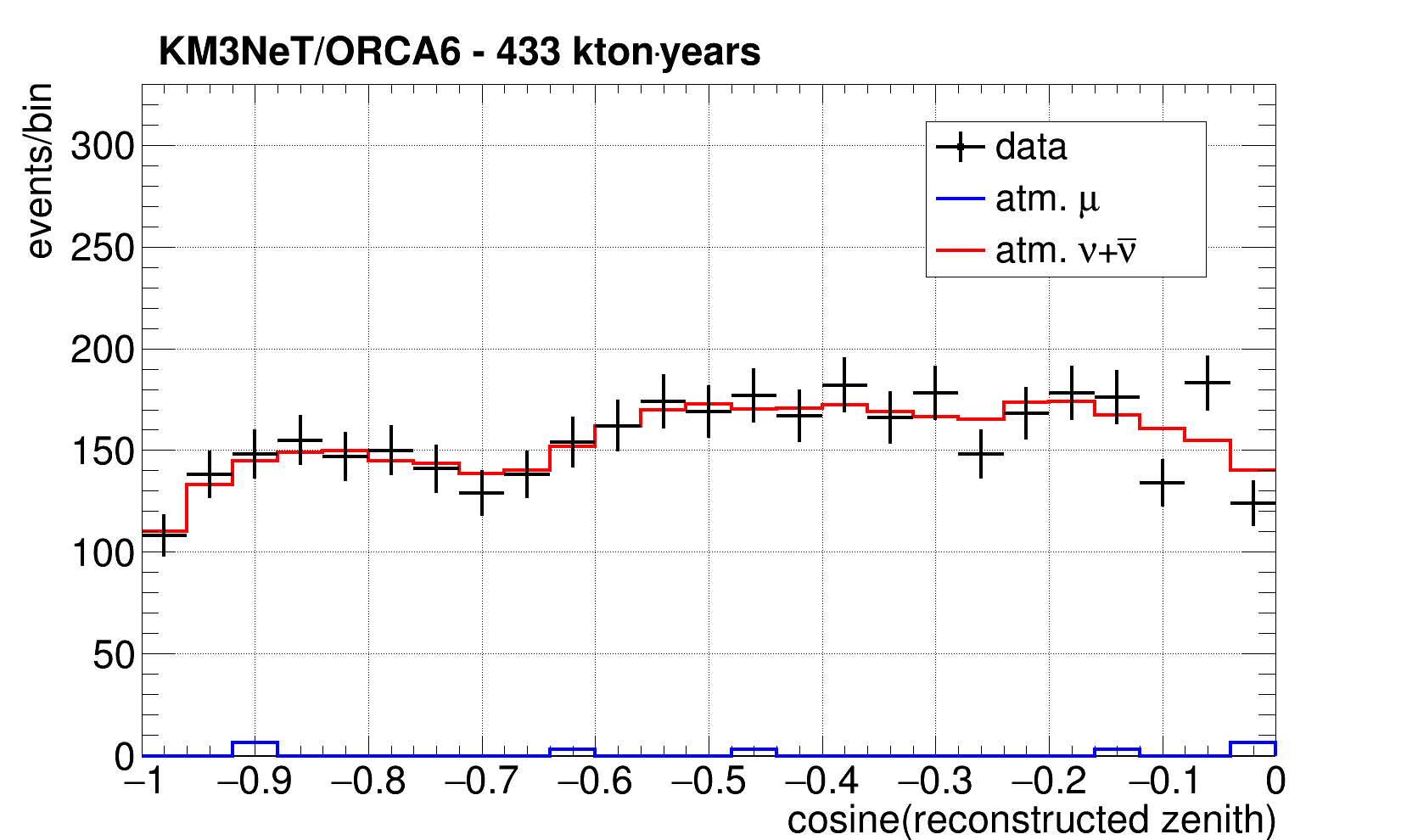}
}
\caption{Distribution of the reconstructed cosine zenith angle for data and MC simulated events after the BDT selection.}
\label{fig:cos_zen_cand}
\end{figure}

\begin{figure}[h]
\centering
\resizebox{0.5\textwidth}{!}{
  \includegraphics{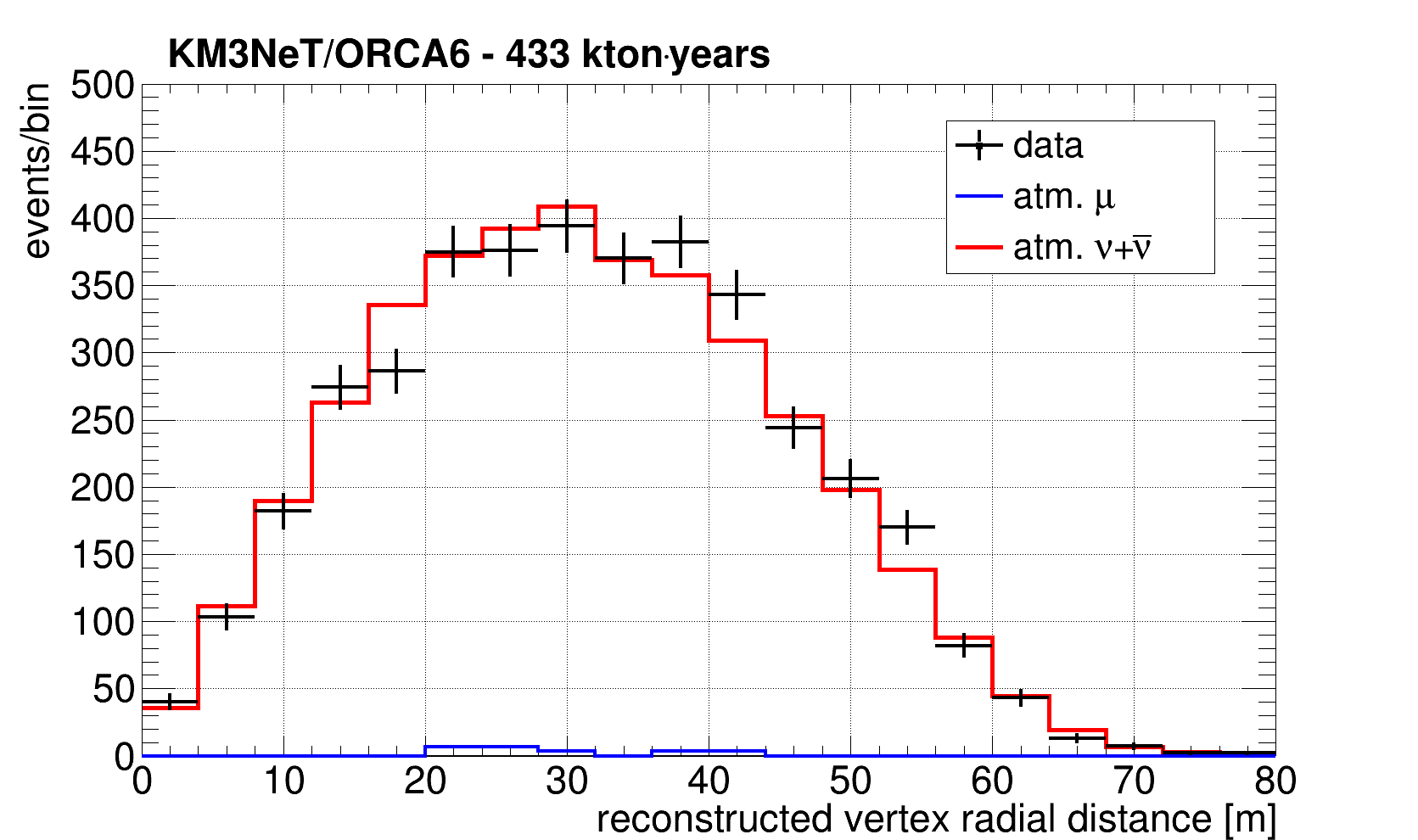}
}
\caption{Distribution of the distance of the reconstructed vertex to the geometrical centre of the instrumented volume for data and MC simulated events after the BDT selection as defined in the text.}
\label{fig:reco_vrtx_r_cand}
\end{figure}

A BDT score cut value is chosen at 0.56 (Fig. \ref{fig:bdt_score_eval_cand}). This results in a high signal efficiency for an almost complete suppression of the atmospheric muon background. The number of selected data and MC simulated events is shown in Table \ref{table:selection}. The contribution of atmospheric muon background is less than $0.6\%$, and $75.5\%$ of the MC simulated events that have been selected are $\nu_{\mu} + \bar{\nu}_{\mu}$ CC events. The distribution of the cosine of reconstructed zenith angle and the distribution of the distance of the reconstructed vertex to the geometrical centre of the instrumented volume is shown in Fig. \ref{fig:cos_zen_cand} and Fig. \ref{fig:reco_vrtx_r_cand} respectively, compared to the MC simulation.

\begin{table}[h]
\resizebox{\columnwidth}{!}{
\begin{tabular}{llll}
\hline\noalign{\smallskip}
& Anti-noise & +Up-going & +BDT \\
\noalign{\smallskip}\hline\noalign{\smallskip}
Data & 2.7$\times 10^8$ & 4.1$\times 10^6$ & 3894 \\
 Atm. $\mu$ & 2.6$\times 10^8$ & 4.4$\times 10^6$ & 23 \\
 Atm. $\nu_e + \bar{\nu}_e$ CC & 2486 & 1590 & 552 \\
 Atm. $\nu_{\mu} + \bar{\nu}_{\mu}$ CC & 15235 & 7673 & 2958 \\
 Atm. $\nu_{\tau} + \bar{\nu}_{\tau}$ CC & 327 & 308 & 162 \\
 Atm. $\nu + \bar{\nu}$ NC & 1279 & 778 &  222 \\
 MC total & 2.6$\times 10^8$ & 4.4$\times 10^6$ & 3917 \\
\noalign{\smallskip}\hline
\end{tabular}
}
\caption{Number of data and MC simulated events at different levels of the selection procedure, corresponding to the 433 kton$\cdot$years exposure.\label{table:selection} }
% Or use
%\vspace*{5cm}  % with the correct table height
\end{table}

\section{Unfolding of the energy spectrum} \label{sec5}

\subsection{Unfolding scheme}
The distribution of a measured quantity in $n$ bins can be expressed as:
\begin{equation} \label{eq:unf_dis}
\begin{split}
y_i = \sum_{j=1}^{m} A_{ij} x_j + b_i  ,   \ \ \ \ \ 1 \leq i \leq n
\end{split}
\end{equation}
where $y_i$ are the components of a vector $\vec{y}$ containing the values of the measured variable and $x_j$ the components of a vector $\vec{x}$, containing the true values of the variable in $m$ bins. The vector $\vec{b}$ accounts for the contribution from all background sources, in each $i$ bin. The matrix $\hat{A}$, called the \textit{response matrix}, allows for corrections to the measurement of a physics variable taking into account possible distortions introduced by the measurement process itself, for example due to the limited size of the detector or its inefficiency. The goal of \textit{unfolding} is to estimate the true spectrum $\vec{x}$ from the measured spectrum $\vec{y}$ using the detector response $\hat{A}$. In this analysis, the energy spectrum of the $\nu_{\mu} + \bar{\nu}_{\mu}$ CC selected events ($\vec{x}$) is unfolded from the reconstructed energy distribution ($\vec{y}$).\par
\begin{figure}[h]
\centering
\resizebox{0.5\textwidth}{!}{
  \includegraphics{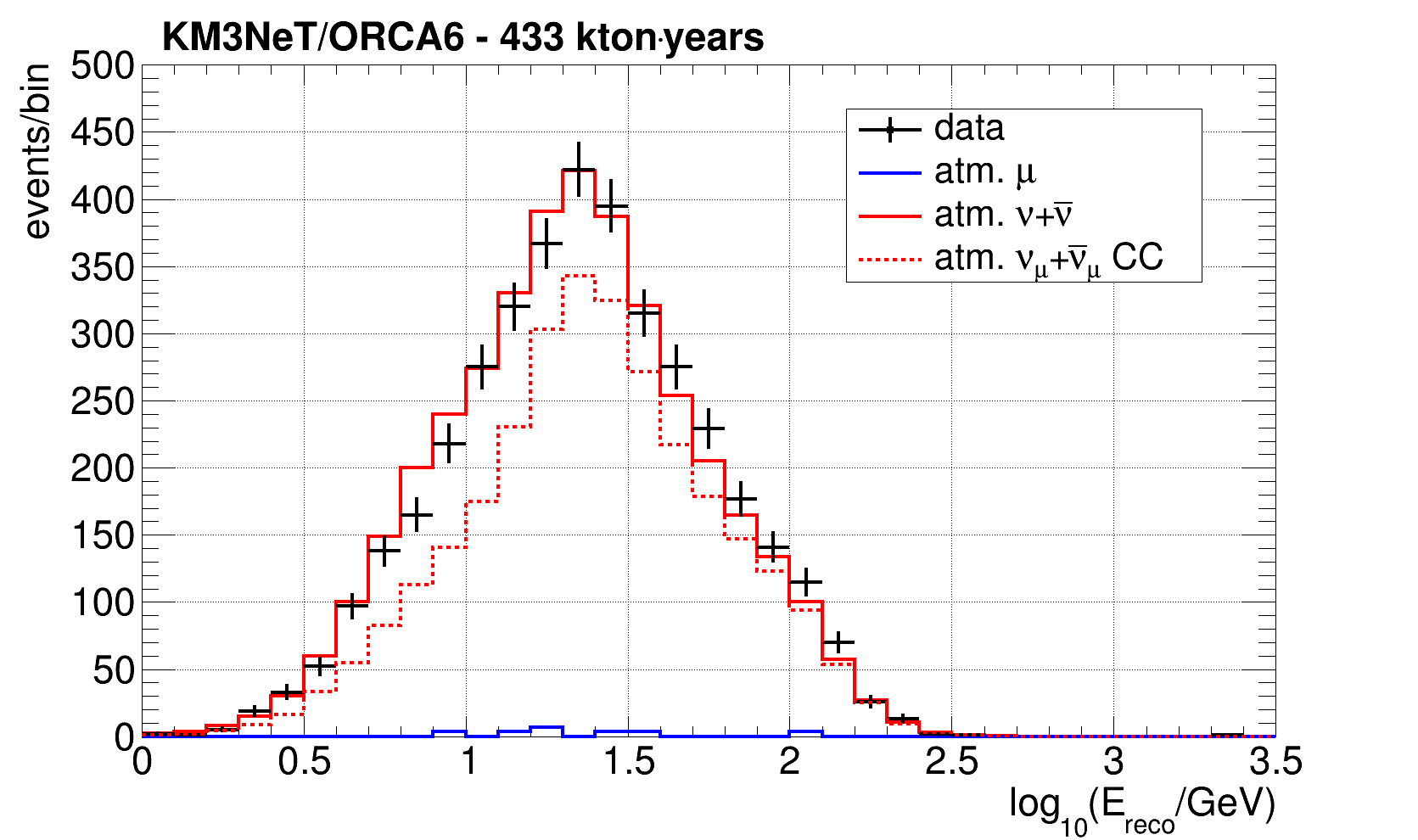}
}
\caption{Distribution of the shower reconstructed energy for data and MC simulated selected events of the event selection.}
\label{fig:JShower_recoE_cand}
\end{figure}

 
The track and shower reconstruction algorithms have been developed and optimised for the completed configuration of KM3NeT/ORCA. Applying these algorithms to the significantly smaller instrumented volume of ORCA6 has an impact on the reconstruction performance. In the case of the energy reconstruction, the performance of the shower reconstruction algorithm is less affected, and therefore the shower reconstructed energy was chosen as the measured distribution $\vec{y}$. This distribution is shown in Fig. \ref{fig:JShower_recoE_cand}.

For the implementation of the unfolding procedure, the TUnfold software is used \cite{tunfold}. The $\vec{x}$ vector in Eq. \ref{eq:unf_dis} is estimated using a least square method with Tikhonov regularisation \cite{tikhonov} and a constraint on the total number of events. TUnfold also allows for a subtraction of background events using the estimation of the MC simulation. The response matrix is constructed using  the $\nu_{\mu} + \bar{\nu}_{\mu}$ CC selected events. The shower-like atmospheric $\nu_e + \bar{\nu}_e$ CC, $\nu_\tau + \bar{\nu}_\tau$ CC, $\nu + \bar{\nu}$ NC, as well as the (negligible, sub-percent) contribution of atmospheric muon events that survive the selection criteria, are treated as background sources within TUnfold. Moreover, considering the limited instrumented volume of ORCA6, the contribution of events with $E_{\nu}>100$ GeV is treated as an overflow to the measured energy spectrum. Αs a result, the flux is measured in the range between 1 GeV and 100 GeV. Three bins are selected for the true neutrino energy, as $\log_{10}(E_{\nu}/\text{GeV}): \{0.0, 0.7, 1.3, 2.0\}$. Details on the unfolding scheme can be found in Appendix \ref{app}. \par
\begin{figure}[h]
\centering
\resizebox{0.5\textwidth}{!}{
  \includegraphics{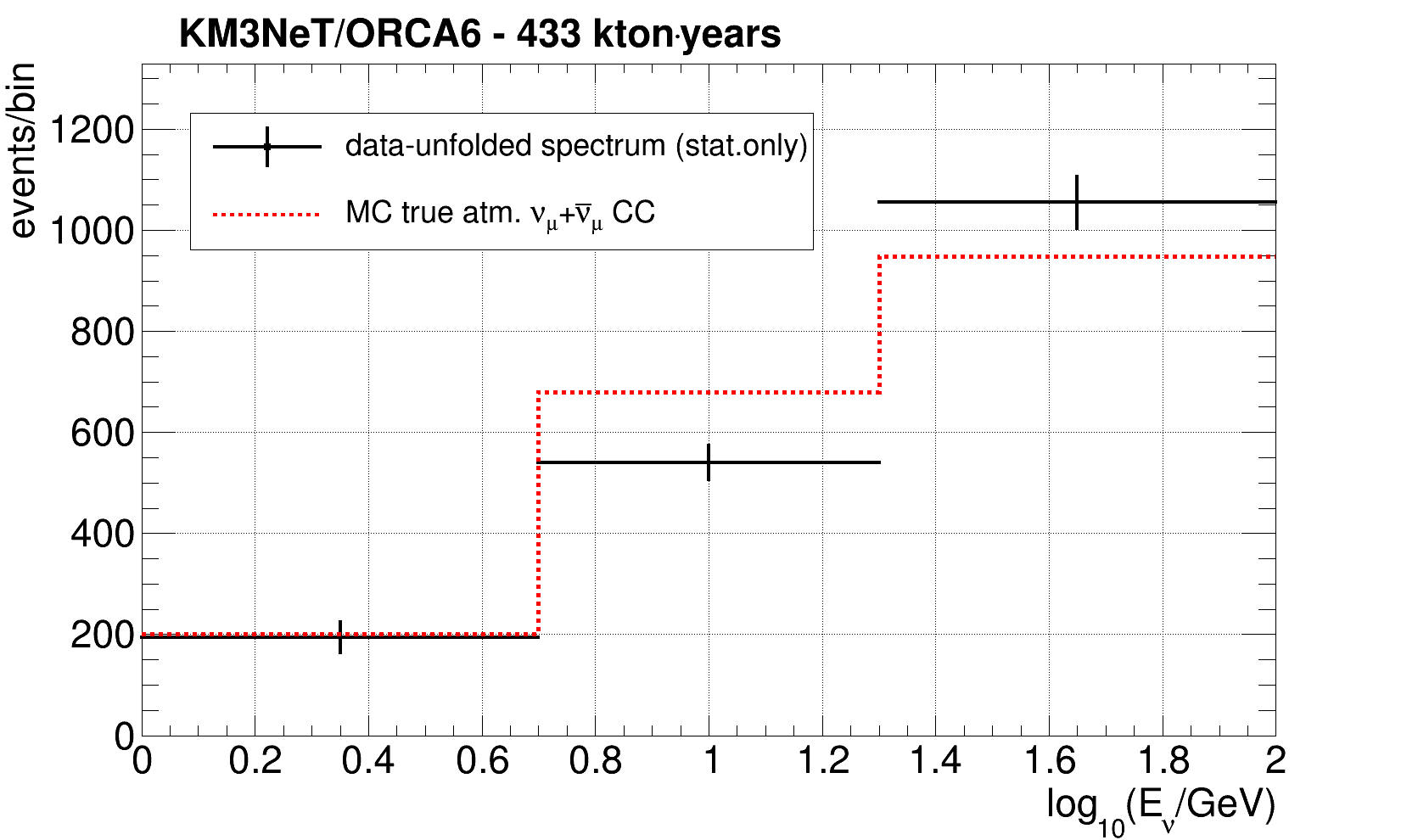}
}
\caption{Distribution of the unfolded energy spectrum. The distribution of the true energy for the $\nu_\mu + \bar{\nu}_\mu$ CC MC simulated events that survive the event selection criteria has been also added as a reference.}
\label{fig:unfolded}
\end{figure}

The unfolded energy spectrum of the $\nu_\mu + \bar{\nu}_\mu$ CC events is shown in Fig. \ref{fig:unfolded}. The results of the unfolding procedure are summarised in Table \ref{tab:unf_tab}.

\begin{table}[h]
\centering
\begin{tabular}{llc}
\hline\noalign{\smallskip}
$\log_{10}(E_{\nu}/\text{GeV})$ & Unfolded & MC $\nu_\mu + \bar{\nu}_\mu $ CC \\
\noalign{\smallskip}\hline\noalign{\smallskip}
 $0.0-0.7$ & $194 \pm  16 \%$ & $202$ \\ 
 $0.7-1.3$ & $540 \pm  6 \%$  & $680$\\
 $1.3-2.0$ & $1055 \pm  5 \%$ & $948$ \\
\noalign{\smallskip}\hline
\end{tabular}
\caption{Number of events for the unfolded energy spectrum with statistical errors. The number of MC simulated events for each bin in true (MC) neutrino energy has been also added as a reference.\label{tab:unf_tab}}
% Or use
%\vspace*{5cm}  % with the correct table height
\end{table}

The differences between the unfolded distribution and the MC distribution are in agreement with what is shown in Fig. \ref{fig:JShower_recoE_cand}. Systematic uncertainties are presented in Section \ref{sec6}, where a discussion of the differences between data and MC simulation is presented. \par

\subsection{Conversion to flux values}

A procedure to convert the unfolded energy spectrum into flux is developed, based on the procedure followed by the Super-Kamiokande experiment \cite{sk}. The atmospheric $\nu_\mu + \bar{\nu}_\mu$ flux for each energy bin $i$ is therefore calculated as:
\begin{equation} \label{eq:flux_calc}
\begin{split}
\Phi_i = \Phi_{\text{MC}}^{\nu_\mu + \bar{\nu}_\mu}(\tilde{E}_i) \cdot \frac{N_i^{\text{unfolded}}}{N_{i,\, \text{MC}}^{\nu_\mu + \bar{\nu}_\mu \, \text{CC}}}
\end{split}
\end{equation}
where $\Phi_i$ is the measured value and $\Phi_{\text{MC}}^{\nu_\mu + \bar{\nu}_\mu}(\tilde{E}_i)$ is the flux predicted by the HKKM14 atmospheric $\nu_\mu + \bar{\nu}_\mu$ flux model \cite{hkkm14}, calculated at the energy weighted bin centre $\tilde{E}_i$ of the selected events, $N_i^{\text{unfolded}}$ is the number of events after unfolding, and $N_{i, \, \text{MC}}^{\nu_\mu + \bar{\nu}_\mu \, \text{CC}}$ is the number of MC simulated $\nu_\mu + \bar{\nu}_\mu \, \text{CC}$ events, as in Tab.~\ref{tab:unf_tab} and in Fig.~\ref{fig:unfolded}.\par 
The reference flux $\Phi_{\text{MC}}^{\nu_\mu + \bar{\nu}_\mu}(E_\nu)$ in Eq.~\ref{eq:flux_calc} is calculated as the zenith and azimuth average of the following expression:
\begin{equation} \label{eq:ref_flux}
\begin{split}
\sum_{\nu_i=\nu_e, \nu_{\mu}}^{} \Big[ \Phi_{\text{MC}}^{\nu_i}(E_\nu,\theta,\phi)\cdot O^{\nu_i\rightarrow\nu_\mu}(E_\nu,\theta) + \\ \Phi_{\text{MC}}^{\bar{\nu}_i}(E_\nu,\theta,\phi)\cdot O^{\bar{\nu}_i\rightarrow\bar{\nu}_\mu}(E_\nu,\theta) \Big]
\end{split}
\end{equation}
where $E_{\nu}$ is the neutrino energy, $\theta$ and $\phi$ are the zenith angle and the azimuthal angle of the neutrino direction respectively, and $O^{\nu_i\rightarrow\nu_j}$ is the oscillation probability of a flavour $i$ to a flavour $j$, assuming azimuthal symmetry. The fluxes $\Phi_{\text{MC}}^{\nu_i}, \ \nu_i=\nu_e,\nu_\mu,\bar{\nu}_e,\bar{\nu}_{\mu}$ are the predicted fluxes from the HKKM14 conventional atmospheric neutrino flux model \cite{hkkm14}, and the neutrino oscillation parameters are the ones from NuFIT 5.2 \cite{nufit}. The flux value for each bin $i$ is calculated with an interpolation of the $\Phi_{\text{MC}}^{\nu_\mu + \bar{\nu}_\mu}$ obtained using Eq. \ref{eq:ref_flux}, at the weighted energy centre $\tilde{E}_i$, and subsequently using the Eq. \ref{eq:flux_calc}.\par

\section{Systematic uncertainties} \label{sec6}
The systematic uncertainties are evaluated by repeating the unfolding procedure described in Section \ref{sec5}, varying input parameters of the MC simulations and taking into account modifications in the response matrix and in the background. This produces modifications to the unfolded flux, which are taken as estimates of the systematic uncertainties following the approach of the ANTARES measurement in \cite{antares2}. The following uncertainties have been considered: \par  

\begin{enumerate}
 
\item The PMT efficiencies and the light absorption length in seawater are modified by $\pm 10\%$ independently, and reffered to as detector response in the following.

\item The effect of uncertainties on the hadronic interaction models used in the simulation of EAS is considered as representative of the uncertainties $\delta$ on the ratio of the neutrino flavours, $(\nu_{\mu}+\bar{\nu}_{\mu})/(\nu_{e}+\bar{\nu}_{e})$, as well as on the ratios of neutrinos and antineutrinos, $\nu_{e}/\bar{\nu}_{e}$ and $\nu_{\mu}/\bar{\nu}_{\mu}$ \cite{barr}. In order to estimate this effect, the corresponding $\Phi_i$, $\Phi_j$ in each ratio are changed as $\Phi_i \rightarrow \Phi_i(1+\delta)$ and $\Phi_j \rightarrow \Phi_j(1- \frac{\Phi_i}{\Phi_j}\delta)$, keeping the total neutrino flux constant.

\item Uncertainties on the neutrino cross section are considered by rescaling the number of $\nu_{\mu}+\bar{\nu}_{\mu}$ CC simulated events with energy dependent factors taken from \cite{nomad}, and $\nu_{\tau}+\bar{\nu}_{\tau}$ CC simulated events by $\pm 20\%$. \par 

\item To evaluate the uncertainties introduced by the unfolding procedure, 2000 pseudo-data sets are generated from the reconstructed energy distribution of Fig. \ref{fig:JShower_recoE_cand}. Pseudo-unfoldings are performed after replacing the measured reconstructed energy distribution with the random generated ones. The associated systematic uncertainties are considered as the $68\%$ quantiles of the residuals between the toy-unfolding results and the nominal result. Additionally, the unfolding is repeated with the model flux modified as $\Phi_{\text{sys}}=\Phi_{\text{nom}} \left( \frac{E}{10 \, \text{GeV}}\right)^{\Delta\gamma}$, where $\Delta \gamma = \pm 0.05$.  \par  

\item A scaling is applied to the MC simulated CC neutrino events with $E_{\text{true}}>500$ GeV, and NC neutrino events with $E_{true}>100$ GeV, to account for the use of different software packages at the light propagation level of MC simulations.

\end{enumerate}
The resulting systematic uncertainties are summarised in Table \ref{tab:sys}. Contributions are considered independent and summed in quadrature.

\begin{table}[h]
\centering
\resizebox{\columnwidth}{!}{
\begin{tabular}{l|ccc}
\hline\noalign{\smallskip}
& \multicolumn{3}{c}{$\log_{10}(E_{\nu}/\text{GeV})$}\\ [0.8ex]
Systematic & $0-0.7$ & $0.7-1.3$ & $1.3-2.0$\\ [0.8ex] 
\noalign{\smallskip}\hline\noalign{\smallskip}
 Detector response & ${}^{+4\%}_{-22\%}$ & ${}^{+19\%}_{-21\%}$ & ${}^{+15\%}_{-10\%}$ \\ [0.8ex]  
 Hadronisation in EAS & ${}^{+14\%}_{-13\%}$  & $\pm 10\%$ & $\pm 9\%$ \\ [0.8ex] 
 Neutrino cross section & $\pm 5\%$  & $\pm 3\%$ & ${}^{+3\%}_{-4\%}$ \\  [0.8ex] 
 Unfolding & ${}^{+21\%}_{-18\%}$ & ${}^{+8\%}_{-7\%}$ & ${}^{+18\%}_{-22\%}$ \\  [0.8ex] 
 Light simulation & $+3\%$  & $-3\%$ & $-13\%$ \\  [0.8ex]
\noalign{\smallskip}\hline
\end{tabular}
}
\caption{Uncertainties for each systematic source category.\label{tab:sys}}
\end{table}

\section{Results} \label{sec7}
The measured flux values and the statistical and systematic uncertainties are presented in Table \ref{tab:meas_stat_sys_tab}. The overall systematic uncertainty for each bin is extracted considering all systematic uncertainty sources presented in Section \ref{sec6} as uncorrelated.

\begin{table}[h]
\centering
\resizebox{\columnwidth}{!}{
\begin{tabular}{llll}
\hline\noalign{\smallskip}
 $\log_{10}(E_{\nu}/\text{GeV})$ & $0.0-0.7$ & $0.7-1.3$ & $1.3-2.0$ \\
\noalign{\smallskip}\hline\noalign{\smallskip}
 $\log_{10}(\tilde{E}_i/\text{GeV})$ & $0.41$ & $0.87$ & $1.50$ \\ [1ex] 
 $\tilde{E}_i^2\Phi_i$ & $1.29\cdot 10^{-2}$ & $4.49\cdot 10^{-3}$ & $1.83\cdot 10^{-3}$  \\ [1ex] 
 \textrm{stat. error} & $\pm 16 \%$ & $\pm 6 \%$ & $\pm 5 \% $ \\ [1ex] 
 \textrm{syst. error} & ${}^{+26\%}_{-32\%}$ & ${}^{+23\%}_{-25\%}$ & ${}^{+25\%}_{-29\%}$ \\
\noalign{\smallskip}\hline
\end{tabular}
}
\caption{Final results on the atmospheric neutrino flux measurement with six DUs of KM3NeT/ORCA. The content of the rows is the following: Energy range, energy weighted bin centre, measured flux multiplied by the squared energy weighted bin center, measured in $\text{GeV}\times \text{s}^{-1}\times \text{sr}^{-1}\times \text{cm}^{-2}$, statistical uncertainty, systematic uncertainty.\label{tab:meas_stat_sys_tab}}
\end{table}

\begin{figure}[h]
\centering
\resizebox{0.5\textwidth}{!}{
  \includegraphics{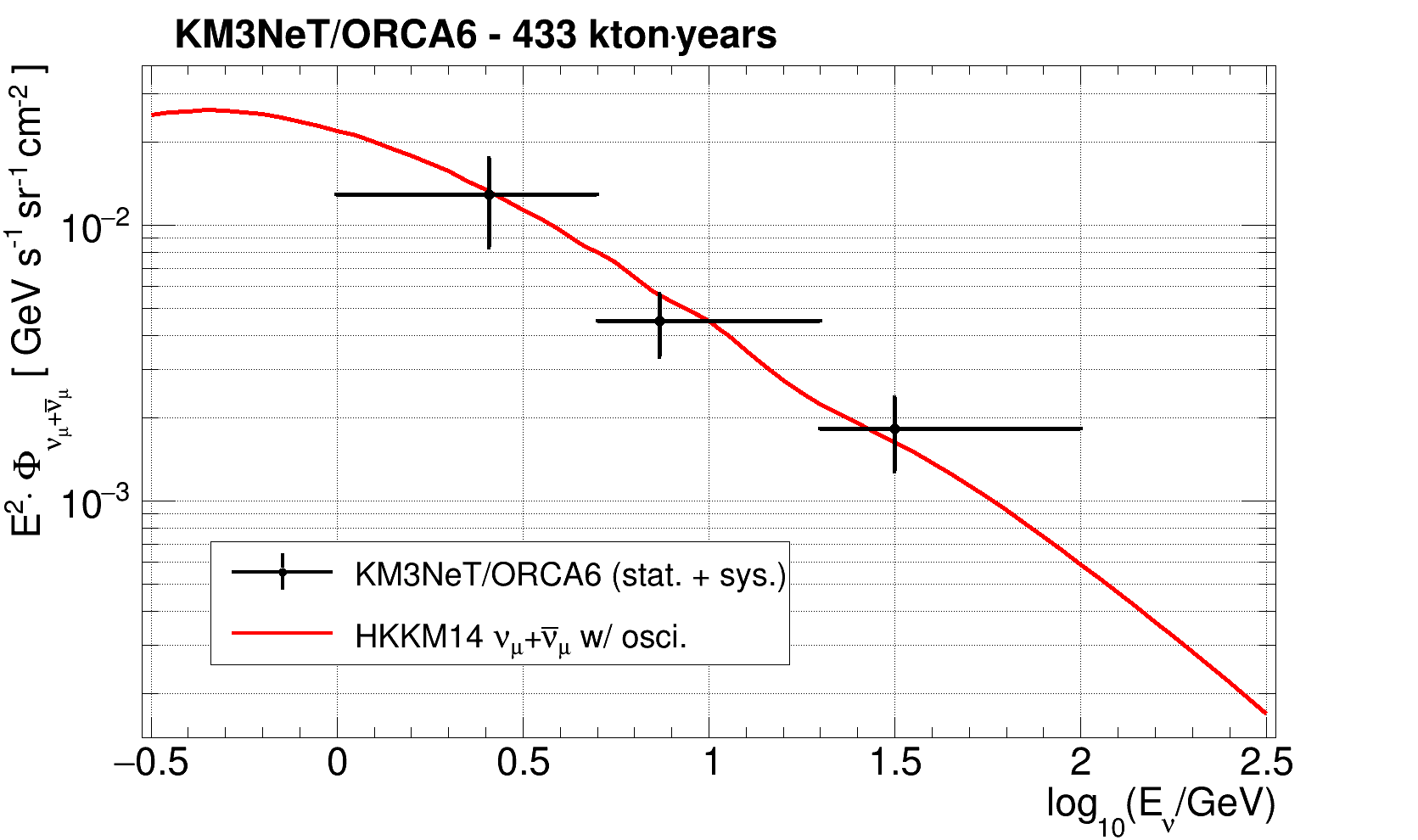}
}
\caption{Atmospheric $\nu_\mu + \bar{\nu}_\mu$ flux measured with ORCA6, multiplied by $\tilde{E}^2$ and superimposed on the atmospheric $\nu_\mu + \bar{\nu}_\mu$ flux predicted by the HKKM14 model.}
\label{fig:measurement_stat_sys_cand}
\end{figure}

The measured atmospheric $\nu_\mu + \bar{\nu}_\mu$ flux is presented in Fig. \ref{fig:measurement_stat_sys_cand}, superimposed on the atmospheric $\nu_\mu + \bar{\nu}_\mu$ flux predicted by the HKKM14 model, according to Eq. \ref{eq:ref_flux}. The error bars represent the quadrature sum of the statistical uncertainty and the overall systematic uncertainty for each bin.

\begin{figure}[h]
\centering
\resizebox{0.5\textwidth}{!}{
  \includegraphics{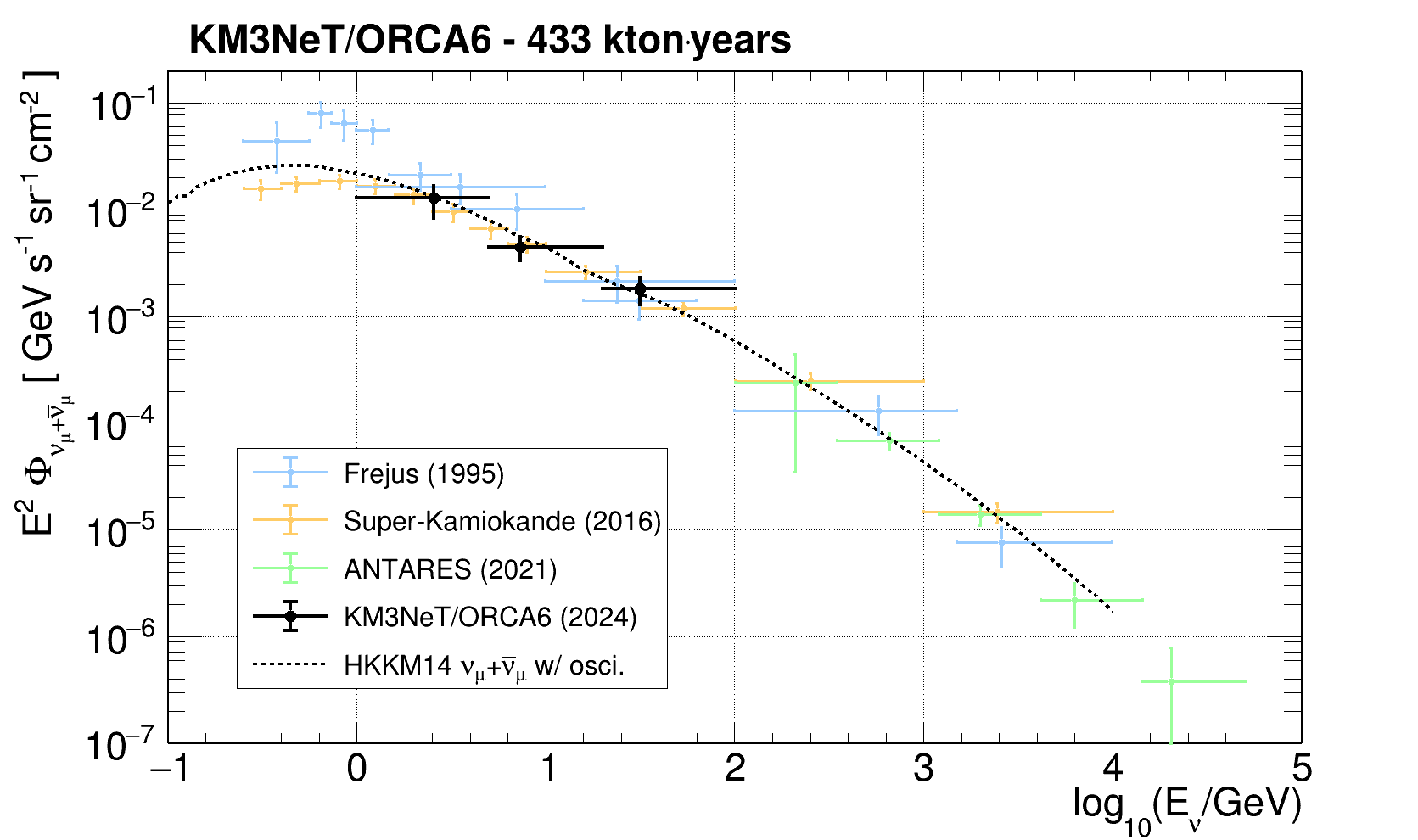}
}
\caption{The atmospheric muon neutrino flux measurement with ORCA6 compared to measurements from other experiments \cite{sk}, \cite{antares2}, \cite{frejus}.}
\label{fig:results_comp_cand}
\end{figure}

The atmospheric $\nu_\mu + \bar{\nu}_\mu$ flux measured with ORCA6 is presented in Fig. \ref{fig:results_comp_cand}, compared to measurements from other experiments, namely from Super-Kamiokande \cite{sk}, ANTARES \cite{antares2}, and Frejus \cite{frejus}. The measurement by Frejus was performed in 1995, before the discovery of neutrino oscillations. For $E_{\nu}>100$ GeV, several measurements have been performed by neutrino telescopes such as AMANDA \cite{amanda}, IceCube \cite{icecube1,icecube2,icecube3}, and ANTARES \cite{antares1}, not depicted in the plot. 

\section{Conclusions} \label{sec8}
The $\nu_\mu + \bar{\nu}_\mu$ flux measured with the ORCA6 configuration is in agreement with the theoretical model, as well as with the experimental results of Super-Kamiokande. For $0.7<\log_{10}(E_{\nu}/\text{GeV})<1.3$, the measured value is about $20\%$ lower with respect to the HKKM14 flux, although within errors. A similar trend is observed in $0.6<\log_{10}(E_{\nu}/\text{GeV})<1.0$ by Super-Kamiokande. This will be further investigated with the ORCA detector, as additional DUs have been deployed and are taking data.\par 


%\begin{figure*}
%\vspace*{5cm}       % Give the correct figure height in cm
%\end{figure*}
%

\section*{Acknowledgements}
The authors acknowledge the financial support of:
%INFRADEV
KM3NeT-INFRADEV2 project, funded by the European Union Horizon Europe Research and Innovation Programme under grant agreement No 101079679;
%Belgium
Funds for Scientific Research (FRS-FNRS), Francqui foundation, BAEF foundation.
%Czeck
Czech Science Foundation (GA\v{C}R 24-12702S);
%France
Agence Nationale de la Recherche (contract ANR-15-CE31-0020), Centre National de la Recherche Scientifique (CNRS), Commission Europ\'eenne (FEDER fund and Marie Curie Program), LabEx UnivEarthS (ANR-10-LABX-0023 and ANR-18-IDEX-0001), Paris \^Ile-de-France Region, Normandy Region (Alpha, Blue-waves and Neptune), France,
%For the CPER
The Provence-Alpes-C\^{o}te d'Azur Delegation for Research and Innovation (DRARI), the Provence-Alpes-C\^{o}te d'Azur region, the Bouches-du-Rh\^{o}ne Departmental Council, the Metropolis of Aix-Marseille Provence and the City of Marseille through the CPER 2021-2027 NEUMED project,
%For IN2P3
The CNRS Institut National de Physique Nucl\'{e}aire et de Physique des Particules (IN2P3);
%Georgia
Shota Rustaveli National Science Foundation of Georgia (SRNSFG, FR-22-13708), Georgia;
%Germany (Max Planck Inst.)
This work is part of the MuSES project which has received funding from the European Research Council (ERC) under the European Union's Horizon 2020 Research and Innovation Programme (grant agreement No 101142396).
%Wurzburg
This work was supported by the European Research Council, ERC Starting grant \emph{MessMapp}, under contract no. $949555$.
%Greece
The General Secretariat of Research and Innovation (GSRI), Greece;
%Italy
Istituto Nazionale di Fisica Nucleare (INFN) and Ministero dell'Unoversit{\`a} e della Ricerca (MUR), through PRIN 2022 program (Grant PANTHEON 2022E2J4RK, Next Generation EU) and PON R\&I program (Avviso n. 424 del 28 febbraio 2018, Progetto PACK-PIR01 00021), Italy; IDMAR project Po-Fesr Sicilian Region az. 1.5.1; A. De Benedittis, W. Idrissi Ibnsalih, M. Bendahman, A. Nayerhoda, G. Papalashvili, I. C. Rea, A. Simonelli have been supported by the Italian Ministero dell'Universit{\`a} e della Ricerca (MUR), Progetto CIR01 00021 (Avviso n. 2595 del 24 dicembre 2019); KM3NeT4RR MUR    Project National Recovery and Resilience Plan (NRRP), Mission 4 Component 2 Investment 3.1, Funded by the European Union -- NextGenerationEU,CUP I57G21000040001, Concession Decree MUR No. n. Prot. 123 del 21/06/2022;
%Morocco
Ministry of Higher Education, Scientific Research and Innovation, Morocco, and the Arab Fund for Economic and Social Development, Kuwait;
%The Netherlands
Nederlandse organisatie voor Wetenschappelijk Onderzoek (NWO), the Netherlands;
%Poland
The grant "AstroCeNT: Particle Astrophysics Science and Technology Centre", carried out within the International Research Agendas programme of the Foundation for Polish Science financed by the European Union under the European Regional Development fund;
The program: 'Excellence initiative-research university' for the AGH University in Krakow; The ARTIQ project: UMO-2021/01/2/ST6/00004 and ARTIQ/ 0004/2021;
%Romania
Ministry of Research, Innovation and Digitalisation, Romania;
%Slovak Republic
Slovak Research and Development Agency under Contract No. APVV-22-0413; Ministry of Education, Research, Development and Youth of the Slovak Republic;
%Spain
MCIN for PID2021-124591NB-C41, -C42, -C43 and PDC2023-145913-I00 funded by MCIN/AEI/ 10.13039/501100011033 and by "ERDF A way of making Europe", for ASFAE/2022/014 and ASFAE/2022 /023 with funding from the EU NextGenerationEU (PRTR-C17.I01) and Generalitat Valenciana, for Grant AST22\_6.2 with funding from Consejer\'{\i}a de  Universidad, Investigaci\'on e Innovaci\'on and Gobierno de Espa\~na and European Union - NextGenerationEU, for \ CSIC-INFRA23013 and for \ CNS2023-144099, Generalitat Valenciana for  \ CIDEGENT/2018/034, /2019/043, /2020/049, /2021/23, for CIDEIG/2023/20, for CIPROM/2023/51 and for GRISOLIAP/2021/192 and EU for MSC/101025085, Spain;
%UAE
Khalifa University internal grants (ESIG-2023-008, RIG-2023-070 and RIG-2024-047), United Arab Emirates;
%UK
The European Union's Horizon 2020 Research and Innovation Programme (ChETEC-INFRA - Project no. 101008324).

\appendix
\renewcommand{\thesection}{\Alph{section}} % Modify numbering format
\setcounter{section}{0}
\renewcommand{\theequation}{A.\arabic{equation}} % Change equation numbering
\setcounter{equation}{0} % Reset equation counter

\section{Unfolding} \label{app}
\begin{figure}[h]
\centering
\resizebox{0.5\textwidth}{!}{
  \includegraphics{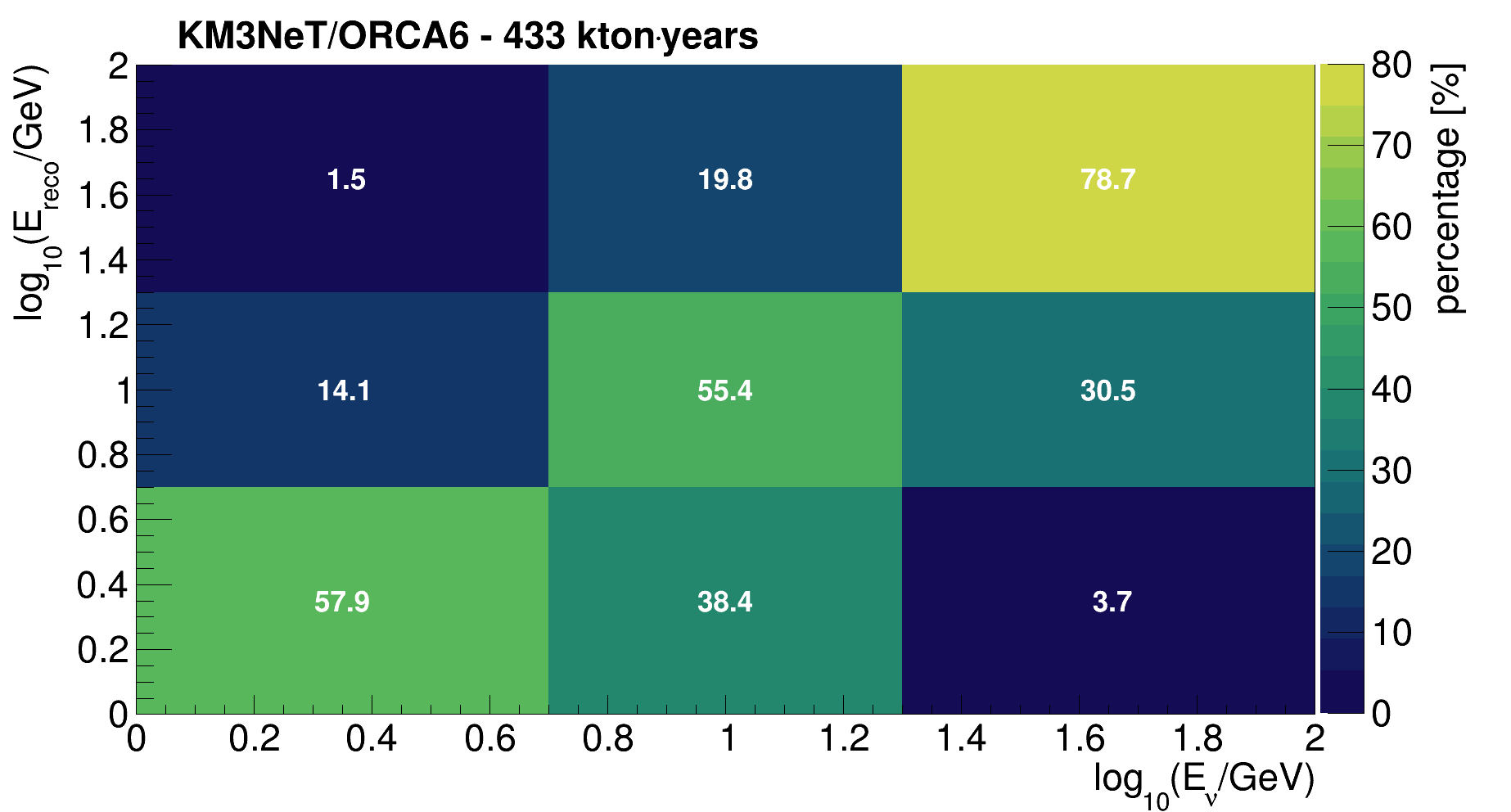}
}
\caption{Response matrix for the $\nu_\mu + \bar{\nu}_\mu$ CC simulated events, normalised to the total number of events in each reconstructed energy bin and expressed as a percentage. The percentage for the diagonal bins indicates the \textit{purity}.}
\label{fig:RespMat_purity_cand}
\end{figure}

The choice of the binning is determined by the bin purity. Three bins for the true neutrino energy are selected as $\log_{10}(E_{\nu}/\text{GeV}): \{0.0, 0.7, 1.3, 2.0\}$. The response matrix is shown in Fig. \ref{fig:RespMat_purity_cand}. In each bin $(i,j)$, the percentage of events with true energy in the $i_{th}$ bin, normalised to the total number of events whose energy is reconstructed in the $j_{th}$ bin is reported. For $i=j$, this percentage indicates the bin purity. \par

The TUnfold algorithm is based on calculating the stationary point of the function $\mathcal{L}$,
\begin{equation} \label{eq:lang}
\begin{split}
\mathcal{L}(\vec{x},\lambda) = \mathcal{L}_1 + \mathcal{L}_2 + \mathcal{L}_3 \, \, ,
\end{split}
\end{equation}
which includes a least square term: 
\begin{equation} \label{eq:lang_1}
\begin{split}
\mathcal{L}_1 = (\vec{y}-\hat{A}\vec{x})^{T}\hat{V}_{yy}^{-1}(\vec{y}-\hat{A}\vec{x})
\end{split}
\end{equation}
where $\hat{V}_{yy}$ is the covariance matrix of the measured distribution, $\vec{y}$. The least square method amplifies statistical fluctuations in the measured distribution and is impacted from the presence of negative correlations between adjacent bins. Hence, a penalty term (referred to as the regularisation term) is added to reduce these effects, which in this analysis has the form: 
\begin{equation} \label{eq:lang_2}
\begin{split}
\mathcal{L}_2 = \tau^2 \, \vec{x}^T \, (\hat{L}^T \, \hat{L}) \, \vec{x} \, \, .
\end{split}
\end{equation}
The regularisation term is expressed in terms of a parameter $\tau$ which defines the regularisation strength. The matrix $\hat{L}$ is selected according to the \textit{derivative} option in TUnfold, i.e. $\hat{L}$ has $m-1$ rows and $m$ columns (where $m$ is the number of bins of the true spectrum), with the non-zero elements of the matrix being:
\begin{equation} \label{eq:lcurve_x}
\begin{split}
L_{i,i}=-1,  \ L_{i,i+1}=+1 \, \, .
\end{split}
\end{equation}
This choice is additionally supported by the fact that migrations between adjacent bins are not negligible, as can be seen in Fig. \ref{fig:RespMat_purity_cand}. A third term is added to ensure that the global normalisation of the measured distribution is preserved:
\begin{equation} \label{eq:lang_3}
\begin{split}
\mathcal{L}_3 = \lambda(Y-e^T\vec{x}) \ \ , \ \ Y=\sum_{i}^{n} y_i \ \ , \ \ e_j=\sum_{i}^{n} A_{ij} \, \, .
\end{split}
\end{equation}

\begin{figure}[h]
\centering
\resizebox{0.5\textwidth}{!}{
  \includegraphics{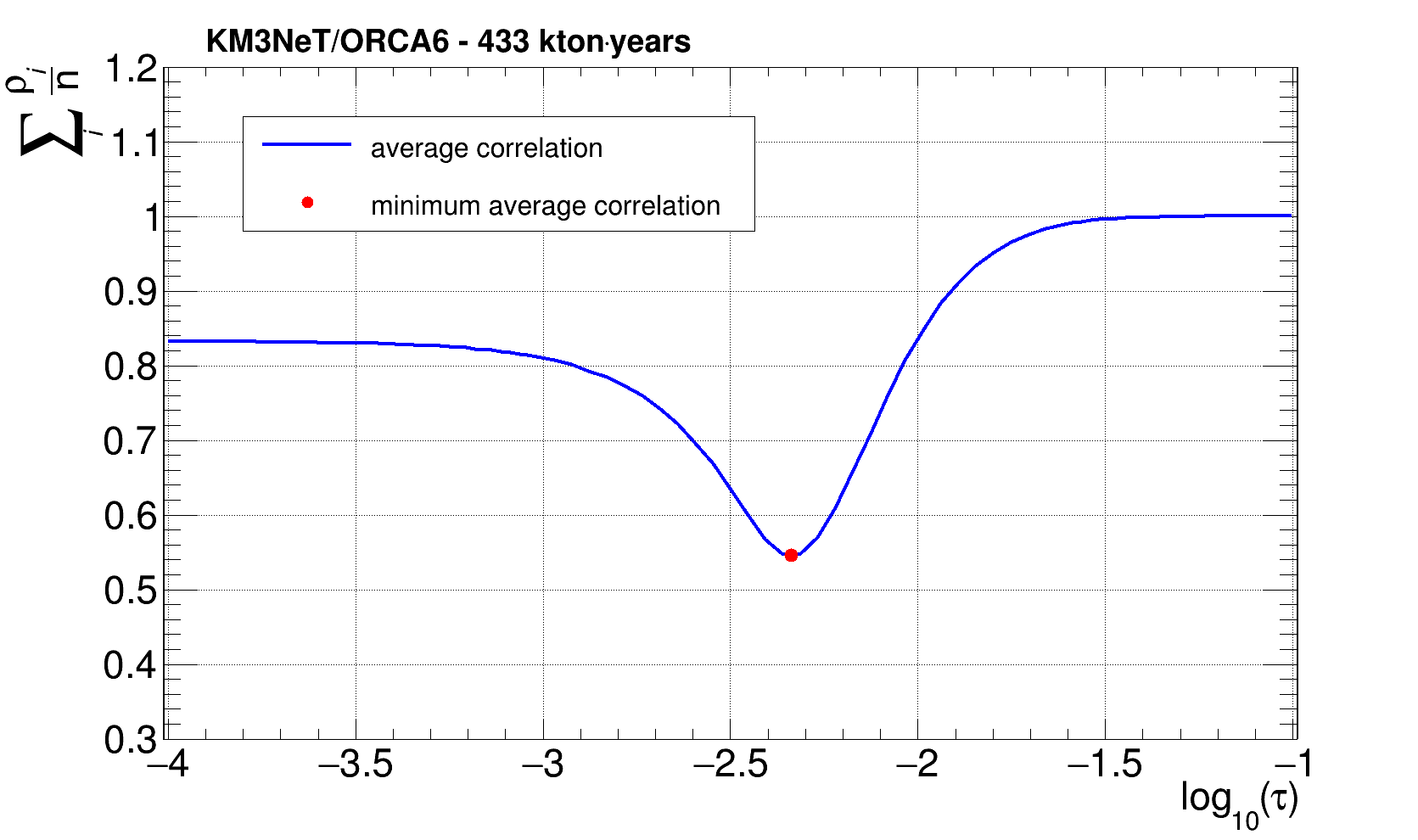}
}
\caption{Average global correlation coefficient for the scaned range of $\tau$. A clear minimum is noticed, indicating the optimal $\tau$ value.}
\label{fig:tauScan_cand}
\end{figure}

The value of the parameter $\tau$ is determined using a method based on the minimisation of the average global correlation coefficient, implemented in TUnfold. The global correlation coefficient is given by:
\begin{equation} \label{eq:gl_rho}
\begin{split}
\rho_i = \sqrt{1-\frac{1}{ (\hat{V}_{xx}^{-1})_{ii}(\hat{V}_{xx})_{ii} }}
\end{split}
\end{equation}
where $\hat{V}_{xx}$ is the covariance matrix of the unfolded spectrum. The choice of the optimal $\tau$, is based on the suppression of strong correlations between the bins of the unfolded spectrum that may be introduced by the unfolding procedure. Hence, the average correlation $\sum_i \rho_i / n$ is used as a metric to be minimised. A total of $100$ unfoldings are performed for $10^{-4}<\tau<10^{-1}$. In Fig. \ref{fig:tauScan_cand}, the average global correlation is shown as a function of the regularisation parameter $\tau$. The value of the regularisation parameter $\tau$ for the minimum average global correlation is $\tau=0.00462$.

The correlation coefficients of the unfolded energy spectrum are $\rho_{12}=-0.232$, $\rho_{23}=+0.113$, $\rho_{13}=-0.185$. No strong correlation is observed between the values of the unfolded spectrum.

%
% BibTeX users please use
% \bibliographystyle{}
% \bibliography{}

\end{document}